\documentclass{article}
\usepackage[utf8]{inputenc} % Handles UTF-8 input encoding
\usepackage[T1]{fontenc}% Font encoding for output (supports more characters)
\usepackage{lmodern} % Latin Modern font (good default, works well with math)
\usepackage[english]{babel} % Language support, hyphenation patterns

\usepackage[margin=1in]{geometry} % Sets 1-inch margins on all sides
\usepackage{enumitem}  % For customizing lists

\usepackage[authoryear, round]{natbib} % Author-year citations
\usepackage[colorlinks=true, linkcolor=blue, citecolor=blue, urlcolor=cyan]{hyperref}

\usepackage{amsmath, amssymb, amsfonts, amsthm}
\usepackage{mathtools} % e.g., \coloneqq
\usepackage{bm} % Bold math symbols

\usepackage{graphicx}
\usepackage{subcaption}
\usepackage{float}
\usepackage{booktabs}
\usepackage{caption}
\usepackage{tikz}
\usepackage{pgfplots}
\pgfplotsset{compat=1.18}
\usetikzlibrary{arrows.meta, patterns, shapes.geometric}

\usepackage[capitalise, nameinlink, noabbrev]{cleveref}

\newcommand{\R}{\mathbb{R}}

\newcommand{\pref}{\succeq}
\newcommand{\spref}{\succ}

\theoremstyle{plain}
\newtheorem{theorem}{Theorem}[section]

\newtheorem{proposition}[theorem]{Proposition}
\newtheorem{corollary}[theorem]{Corollary}

\theoremstyle{definition}
\newtheorem{definition}[theorem]{Definition}

\newtheorem{example}{Example}

\theoremstyle{remark}
\newtheorem{remark}{Remark}

\title{\LARGE\bf Beyond Scalars: Zonotope-Valued Utility for Representation of Multidimensional Incomplete Preferences **(Incomplete Draft)**}

\author{
    Behrooz Moosavi Ramezanzadeh\thanks{University of Pittsburgh, Pittsburgh, PA, USA. \texttt{behroozmoosavi@pitt.edu}}
  \and
    Arie Beresteanu\thanks{University of Pittsburgh, Pittsburgh, PA, USA. \texttt{arie@pitt.edu}}
}

\date{\today} % leave empty to suppress date; or use \date{\today}

\begin{document}
\maketitle

\begin{abstract}
\normalsize
In this paper, we propose a new framework for representing multidimensional incomplete preferences through zonotope-valued utilities. Deviations from traditional approaches to $\mathbb{R}^1$-valued utility representation of preferences, include interval-valued utilities and vector-valued utilities. These extensions, however, fail to capture the complexity of preferences where alternatives remain incomparable due to conflicting criteria across multiple dimensions. Our method maps each alternative to a zonotope, a convex geometric object in \(\mathbb{R}^m\) formed by weighted Minkowski sums of intervals, which encapsulates the multidimensional structure of preferences. The set-valued nature of these payoffs stems from multiple sources, including non-probabilistic uncertainty, such as imprecise utility evaluation due to incomplete information about criteria weights, and probabilistic uncertainty arising from stochastic decision environments. We provide an axiomatization that defines preference as one alternative's zonotope differing from another's within the non-negative orthant of \(\mathbb{R}^m\). This framework generalizes existing representations and provides a visually intuitive and theoretically robust tool for modeling trade-offs among each dimension, while preferences are incomparable. 
\end{abstract}

\vspace{0.5cm}
 {\large \textbf{Keywords}: Decision theory, Interval Orders, Zonotopes, Minkowski Sum, Incomplete Preferences.

\section{Introduction}
\label{sec:introduction}

Central to having an $\mathbb{R}^1$-valued utility representation of preferences is the assumption that decision makers have a complete order over the set of alternatives. Albeit still ubiquitous in applied and empirical fields, completeness has been challenged by decision- theorists since \cite{aumann-1962}. A large body of literature on decision theory offers alternatives to completeness. Early literature on incomplete preferences remains one-dimensional. The decision maker employs a single decision criterion, but one that can not completely rank the set of alternatives. As a result, instead of preferences being represented by a single utility value, \cite{fishburn-1970} provides a system of axioms that relaxes completeness and under which preferences can be represented as an interval utility. Similarly, \cite{Bewley-1986}) considers imprecise uncertainty and shows that expected utility can be represented as an interval. 

A single criterion for ranking alternatives often fails to capture the multidimensional aspect of choice where decision makers have conflicting criteria across multiple dimensions, such as economic cost, quality, or ethical considerations. Although each dimension of choice by itself may offer a complete order of the alternatives, using multidimensional criteria can yield an incomplete order (\cite{ok-2002}). This paper offers a general approach to modeling incomplete preferences that (i) takes into account the possibility that decision involves multiple criteria and (ii) that each dimension by itself does not necessarily provide a complete order. 
 
\paragraph{Literature Review}
The study of incomplete preferences has been a significant focus in decision theory, challenging the assumption that all alternatives can be ranked. \citet{aumann1962utility} demonstrated that realistic decision scenarios often involve incomparability, particularly under uncertainty, necessitating models that relax the completeness axiom. Building on this, \citet{ok2002utility} developed a representation of incomplete preferences using sets of scalar utility functions, where an alternative \( x \) is preferred to \( y \) if \( u(x) \geq u(y) \) for all utility functions \( u \) in the set. Similarly, \citet{dubra2002model} proposed multi-utility representations to model preferences in the presence of risk, and they established conditions for representing indecisiveness. These frameworks provide a robust foundation for handling incomplete preferences but rely on scalar utilities, limiting their capacity to address multidimensional trade-offs comprehensively.
\par
To model incomparability more effectively, \citet{fishburn1970utility,fishburn1985interval} introduced interval orders, representing preferences as intervals on the real line, where \( x \succeq_i y \) if the interval for \( x \) lies entirely to the right of \( y \). Characterized by the absence of the ``2+2'' subposet, interval orders offer a numerical approach to preferences with ranges, a concept foundational to our work. \citet{evren2011multi} extended multi-utility representations to continuous settings, enhancing their applicability to complex economic models. Additionally, \citet{kreinovich2014decision} and \citet{denoeux2020interval} explored set-valued utilities to address uncertainty, associating alternatives with sets to account for imprecise evaluations. However, these approaches focus primarily on uncertainty rather than the structural incomparability central to multidimensional preferences, highlighting the need for a more geometric representation.
\par

Building on this foundation, \citet{dubra2002model} develop an expected utility theory that does not require the completeness axiom, allowing for incomplete preferences in decision-making under uncertainty. Their approach provides a more general framework that can handle situations where individuals may be indifferent or unable to compare certain outcomes, which is particularly relevant for modeling real-world decision processes where uncertainty complicates preference formation.

Further advancing the representation of incomplete preferences, \citet{Cerreia-Vioglio2016} introduce a lexicographic expected utility framework that accommodates both incomplete and non-Archimedean preferences. By relaxing the Archimedean axiom, their model allows for preferences that may not be continuous, which is useful for capturing situations where small changes in outcomes do not affect preferences. This approach provides an alternative geometric representation that complements the zonotope-valued utilities proposed in this paper, as it explicitly incorporates multidimensionality through lexicographic orders.

Traditional multi-attribute utility theory, as developed by \citet{keeney1976decisions}, aggregates multiple criteria into a single scalar value, often oversimplifying the trade-offs inherent in multidimensional decisions. For instance, combining economic and qualitative factors into one number may obscure critical distinctions in preference structures. Vector-valued utilities, as explored by \citet{dubra2002model}, provide a multidimensional perspective but lack the geometric structure needed to fully represent complex preferences. In contrast, our approach utilizes a geometric object, zonotopes, as described by \citet{ziegler1995lectures}, to model preferences as compact convex sets formed by Minkowski sums of intervals. Unlike models addressing probabilistic uncertainty \citep{walley1991statistical} or risk \citep{karni2015continuity}, our framework emphasizes the geometric structure of preference relations, synthesizing insights from interval orders and set-valued utilities to offer a new, more intuitive representation of multidimensional preferences without the completeness axiom.

\paragraph{Contribution:}
This paper introduces a set of axioms that can be represented by a zonotope-valued utility as a framework for modeling multidimensional incomplete preferences, addressing limitations of both scalar and vector-based models. By decomposing preference relations into interval orders and mapping each alternative to a zonotope in \(\mathbb{R}^m\), we establish a rigorous axiomatization ensuring \( x \succeq y \iff U(x) \ominus_e U(y) \subseteq \mathbb{R}^m_{\ge 0} \) (see \citet{beresteanu2025extendedsetdifference} for more details on extended set difference), with the minimal dimension \( m \) equal to the preference's interval dimension. The set-valued nature of our payoffs arises from multiple sources, including non-probabilistic uncertainty, such as imprecise evaluation of utility due to incomplete information about criteria weights \citep{kreinovich2014decision}, and probabilistic uncertainty stemming from stochastic outcomes in decision environments \citep{walley1991statistical}. This approach generalizes existing representations while revealing the geometry of trade-offs through zonotope shapes, providing a visually intuitive and theoretically robust tool for economic modeling and multi-criteria decision analysis. By capturing both structural incomparability and uncertainty-driven set-valued payoffs, our framework offers significant potential for advancing theoretical research and practical applications in decision theory.

\section{Preliminaries and Background}\label{sec:preliminaries}
To establish the theoretical foundation for representing incomplete preference relations with set-valued utilities, we begin by reviewing key concepts and results from order theory and utility representation. This section introduces the essential elements of preorders and set-theoretic structures, reviewing the classical definitions of partially ordered sets and their properties. We then explore the notions of width and linear extensions, which characterize the complexity of incomparability and provide a basis for multi-dimensional orderings, including product orders which are critical to our framework. Finally, we summarize the theory of interval orders, which illustrate the geometric representation of preferences. These concepts form the building blocks for our subsequent analysis of set-valued utilities.

% Additional overview paragraph for coherence
In what follows, in order to make the paper self-contained, each definition and theorem is accompanied by intuitive commentary and examples to guide the reader. We emphasize how these classical results will later support the decomposition of an arbitrary incomplete preference into more tractable components.

\subsection{Basic Definitions}
\label{subsec:basicdef}

We study preference relations over a set $X$ of mutually exclusive and exhaustive alternatives, such as a collection of consumer goods or potential actions. We start with the following definitions, drawn from the literature on order theory which provide the structure needed to analyze comparability and incomparability in multidimensional settings. 

A natural starting point is the concept of a preorder, which formalizes how alternatives are ranked, allowing for both strict preferences and for indifference \citep[Section 1.1]{davey-2002}.

\begin{definition}[Preorder and Poset]\label{def:preorder-poset}
A binary relation $\succeq$ on $X$ is a \emph{preorder} if it is:
\begin{itemize}
 \item \textbf{Reflexive}: $x\succeq x$ for all $x\in X$;
 \item \textbf{Transitive}: if $x\succeq y$ and $y\succeq z$, then $x\succeq z$.
\end{itemize}
If, in addition, 
\begin{itemize}
    \item \textbf{Antisymmetric:} $x\succeq y$ and $y\succeq x$ imply $x=y$.   
\end{itemize}
$(X,\succeq)$ is called a \emph{partially ordered set (poset)} \citep[Chapter~1]{davey-2002}.
\end{definition}

\begin{remark}
Preorders admit \emph{indifference} among distinct elements, while posets collapse such ties into equivalence classes. To pass from a preorder to a poset, one forms the quotient defined in Definition~\ref{def:quotient-poset}. 
\end{remark}

% Do we really need such a trivial example in an economic theory journal?
\begin{example}
Let $X=\{a,b,c\}$ with $a\succeq b$, $b\succeq c$, and $c\succeq a$. Then $\succeq$ is reflexive and transitive but not antisymmetric, so $\succeq$ is a preorder but $(X,\succeq)$ is not a poset.
\end{example}

% Added transition to next subsection
This fundamental distinction between preorders and posets arises repeatedly as we address the treatment of indifferences and shift our attention to strict preference structures in subsequent sections.

\subsection{Comparability: Chains, Antichains, and Quotients}
Understanding which elements are comparable versus incomparable is central to measuring the complexity of a poset.

\begin{definition}[Total Order and Chain]\label{def:total-order}
A preorder $\succeq$ on $X$ is a \emph{total order} if for every pair $x,y\in X$, either $x\succeq y$ or $y\succeq x$. Such a poset is also called a \emph{chain} \citep[Chapter~1]{davey-2002}.
\end{definition}

\begin{definition}[Antichain and Width]\label{def:antichain-width}
An \emph{antichain} in a poset $(X,\succeq)$ is a subset $A\subseteq X$ in which no two distinct elements are comparable. The \emph{width} $w(X,\succeq)$ is the maximum cardinality of an antichain \citep[Section~3.2]{davey-2002}.
\end{definition}

\begin{example}
In the poset on $\{1,2,3,4\}$ with only $1\succeq2$ and $3\succeq4$, the sets $\{1,2\}$ and $\{3,4\}$ are chains, while $\{2,3\}$ is an example for maximal antichain. Hence the width is~2.
\end{example}

\begin{definition}[Quotient Poset]\label{def:quotient-poset}
Given any preorder $\succeq$ on $X$, define an equivalence relation by $x\sim y$ iff $x\succeq y$ and $y\succeq x$. For any $x \in X$, let $[x] = \{y \in X : x \sim y \}$. The \emph{quotient poset} $(X/\sim,\succeq')$ has elements $[x]$ and
\[
 [x]\succeq'[y] \iff x\succeq y.
\]
The quotient set is a poset since $\succeq'$ satisfies reflexivity, transitivity and antisymmetry \citep[Section~1.3]{davey-2002}.
\end{definition}

% Added explanatory remark
The width of the quotient poset gives insight into the maximum degree of true incomparability after collapsing indifference; this will inform later decomposition results via Dilworth's theorem.

\subsection{Resolving Incomparabilities: Linear Extensions and Dimension}
To embed incomplete preferences into a numeric utility framework, one often constructs linear extensions.

\begin{definition}[Linear Extension]\label{def:linear-extension}
A relation $R$ on $X$ is a \emph{linear extension} of a preorder $\succeq$ if:
\begin{enumerate}
 \item $R$ is a total preorder;
 \item $x\succeq y$ implies $x R y$;
 \item if $x \succ y$ (i.e. $x\succeq y$ but not $y\succeq x$), then $x R y$ and not $y R x$.
\end{enumerate}
Equivalently, $R$ is a total order on the quotient $X/\sim$ respecting the original comparabilities \citep[Section~1.3]{davey-2002}.
\end{definition}

\begin{theorem}[Szpilrajn's Extension Theorem, 1930]\label{thm:szpilrajn}
Every poset can be extended to at least one linear order. That is, any preorder $\succeq$ admits a linear extension $R$ \citep{Szpilrajn1930}.
\end{theorem}

\begin{definition}[Dimension of a Poset]\label{def:dimension}
The \emph{dimension} of a poset $(X,\succeq)$, denoted $\dim(X,\succeq)$, is the minimum number of linear extensions whose intersection recovers $\succeq$:
\[
  \succeq = \bigcap_{i=1}^k R_i,
\]
with each $R_i$ a total order.  One always has $\dim(X,\succeq)\le|X|$ for finite $X$ \citep[Section~3.3]{davey-2002}.
\end{definition}

\begin{theorem}[Dushnik–Miller, 1941]\label{thm:dushnik-miller}
A poset is the intersection of all its linear extensions, and this intersection is the original relation. Furthermore, each linear extension can be represented by a real-valued utility function $u_R$ with $xRy \iff u_R(x)\ge u_R(y)$ \citep{dushnik-1941}.
\end{theorem}

\begin{theorem}[Dilworth, 1950]\label{thm:dilworth}
In any finite poset, the width equals the minimum number of chains needed to cover the set:  $w(X,\succeq)=\min\{k:X=C_1\cup\cdots\cup C_k\}$, where each $C_i$ is a chain.  Equivalently, $\succeq$ is the intersection of $w(X,\succeq)$ linear extensions \citep{dilworth-1950}.
\end{theorem}

% Added linkage to interval orders
These results set the stage for replacing linear extensions with richer geometric constructs—interval orders—which allow us to capture preference gaps more finely.

\subsection{Interval Orders and Interval Dimension}
Interval orders introduce a geometric viewpoint via interval representations.

\begin{definition}[Interval Order]\label{def:interval-order}
A preorder $\succeq_i$ on $X$ is an \emph{interval order} if it satisfies:
\begin{itemize}
 \item \textbf{Ferrers Property}: $x\succ_i y$ and $z\succ_i w$ imply $x\succeq_i w$ or $z\succeq_i y$;
 \item \textbf{Semitransitivity}: if $x\succ_i y\succ_i z$, then $x$ outranks all that $y$ outranks, and all outranking $z$ also outrank $y$.
\end{itemize}
Equivalently, there is an assignment of closed intervals $[\underline u(x),\overline u(x)]$ in $\R$ such that
\[
 x\succeq_i y \iff \underline u(x) \ge \overline u(y).
\]
This characterization appears in \citep{fishburn-1970} and is discussed in depth in \citep{fishburn-1985}.
\end{definition}

\begin{definition}[Forbidden $2+2$ Subposet]\label{def:forbidden-2plus2}
A poset contains a \emph{$2+2$} if it has distinct elements $a_1,a_2,b_1,b_2$ with $a_1\succ a_2$, $b_1\succ b_2$, and no other relations among them.  A relation is an interval order if and only if it avoids any 2+2 pattern \citep[Chapter~2]{fishburn-1970}.
\end{definition}

\begin{definition}[Interval Dimension]\label{def:interval-dimension}
The \emph{interval dimension} of $(X,\succeq)$ is the smallest integer $k$ such that
\[
 \succeq = \bigcap_{i=1}^k \succeq_i,
\]
with each $\succeq_i$ an interval order.  As with poset dimension, one has $\dim_{\mathrm{int}}(X,\succeq)\le|X|$ \citep[Chapter~2]{fishburn-1985}.
\end{definition}

% Added remark on how interval dimension bridges to zonotopal representation
By viewing each interval order as a projection of a higher-dimensional embedding, we prepare the ground for the zonotope‐based representations introduced in Section~3.

\subsection{Product Orders}
To model multi-criteria preferences, we use product orders on Cartesian products of attribute sets.

\begin{definition}[Product Order]\label{def:product-order}
Let $(X,\succeq_X)$ and $(Y,\succeq_Y)$ be posets. The \emph{product order} $\succeq_{X\times Y}$ on $X\times Y$ is defined by
\[
 (x_1,y_1) \succeq_{X\times Y} (x_2,y_2) \iff x_1\succeq_X x_2 \text{ and } y_1\succeq_Y y_2.
\]
This extends naturally to finite Cartesian powers $X^m$, producing multidimensional orders \citep{birkhoff-1940}.
\end{definition}

\begin{remark}
Product orders capture the idea of \emph{Pareto} or \emph{componentwise} comparison and are essential when we represent preferences by set-valued utilities in $\R^m$.
\end{remark}

These definitions and classical theorems establish the toolkit we use in Section ~\ref{sec:zonotope_representation} to decompose an arbitrary incomplete preference into a collection of interval order components. They also provide the standard references for readers seeking deeper proofs and further examples.

\section{Modeling Incomplete Preferences via Zonotope-Valued Utilities}
\label{sec:zonotope_representation}
Let $X$ be a set of alternatives, and let $\pref$ be a preorder on $X$, interpreted as a preference relation ("at least as good as"). We introduces an order-theoretic axiomatization on $(X, \pref)$ that guarantees the existence of a set-valued utility function $U: X \to \mathcal{K}(\mathbb{R}^n)$ for an integer $n$ specified in subsection \ref{subsec:countable_case}. Here, $\mathcal{K}(\mathbb{R}^n)$ denotes the family of non-empty compact convex subsets of $\mathbb{R}^n$. A distinguishing feature of the proposed representation is that each utility set $U(x)$ is a zonotope.\footnote{A zonotope in $\mathbb{R}^n$ is a convex polytope that can be constructed as the Minkowski sum of a finite set of line segments.} %The dimension $n$ of the Euclidean space can be chosen, typically $n=m$ where $m$ is the number of interval orders in the decomposition of the preference relation.

%The following paragraph doesn't make any sense placed here:
%\paragraph{States and probability vectors.}
%A \emph{state} labels any contingency—time period, scenario, stakeholder, or region—along which outcomes are assessed. 
%For each attribute $k$ we associate a non‑negative probability vector $\mathbf p_k=(p_{k1},\dots,p_{kd})$ with $\sum_{s=1}^{d} p_{ks}=1$. 
%The coordinate $p_{ks}$ represents the share of attribute $k$’s interval utility that is relevant when state $s$ occurs; geometrically $\mathbf p_k$ is the direction in $\mathbb{R}^{d}$ along which the segment $I_k(x)$ is extruded, while economically it functions as a state‑specific weight or discount factor. 
%Different vectors therefore tilt the generators $v_k$ in different directions, producing the cross‑state trade‑offs encoded in each utility zonotope.

% you can put it in somewhere else when we discuss the prob
\subsection{The Countable Case}
\label{subsec:countable_case}

\subsubsection{Axioms}
\label{subsubsec:axioms_countable}

Let \( X \) be a countable set and \( \pref \) a preorder on \( X \) (reflexive and transitive). We impose the following axioms to ensure the existence of a zonotope-valued utility function:

\begin{enumerate}[label=(A\arabic*)]
 \item \label{axiom:interval_decomposition} \emph{Interval-Order Decomposition}: The preference relation \( \pref \) can be expressed as \( \pref = \bigcap_{k=1}^m \pref_k \), where each \( \pref_k \) is an interval order on \( X \), and \( m \) is finite.
 \item \label{axiom:non_degeneracy} \emph{Non-Degeneracy}: If \( x \) and \( y \) are incomparable under \( \pref \) (neither \( x \pref y \) nor \( y \pref x \)), then \( x \) and \( y \) are incomparable in at least one \( \pref_k \).
 \item \label{axiom:strict_consistency} \emph{Strict Preference Consistency}: If \( x \spref y \) (i.e., \( x \pref y \) and not \( y \pref x \)), then \( x \spref_k y \) for some \( k \).
 \item \label{axiom:vector_independence}
      \emph{Vector‑Independence (statewise Sure‑Thing)}:  
      For every non‑empty set of states $S\subseteq\{1,\dots,d\}$ and
      for all $x,y,z,t\in X$ satisfying
      \[
        x\!\restriction_{S^{c}} = y\!\restriction_{S^{c}},\quad
        z\!\restriction_{S^{c}} = t\!\restriction_{S^{c}},\quad
        x\!\restriction_{S} = z\!\restriction_{S},\quad
        y\!\restriction_{S} = t\!\restriction_{S},
      \]
      we have
      \[
        x \pref y \;\Longleftrightarrow\; z \pref t .
      \]

\end{enumerate}

Axiom \ref{axiom:interval_decomposition} ensures that the preference relation can be decomposed into a finite number of interval orders, providing a structured foundation for the utility representation. Axiom \ref{axiom:non_degeneracy} preserves the logical structure of incomparability, while Axiom \ref{axiom:strict_consistency} ensures consistency in strict preferences across component orders. Axiom \ref{axiom:separability} facilitates numerical representation by allowing interpolation between strictly preferred elements.

\subsubsection{Existence Theorem}
\label{subsubsec:existence_countable}

Each interval order \( \pref_k \) admits real-valued functions \( \underline{u}_k, \overline{u}_k: X \to \mathbb{R} \) such that:
\[
x \pref_k y \iff \underline{u}_k(x) \ge \overline{u}_k(y), \quad \underline{u}_k(x) \le \overline{u}_k(x).
\]
Define the interval \( I_k(x) = [\underline{u}_k(x), \overline{u}_k(x)] \subset \mathbb{R} \).
Formally, we define the zonotope-valued utility function as follows. Given intervals $I_k(x)$ corresponding to each interval order, and a set of vectors (basis) $v_1,v_2,\dots,v_m\in\mathbb{R}^n$ (typically $n=m$), define:
\[
U(x)=\bigoplus_{k=1}^m I_k(x)v_k = \left\{\sum_{k=1}^m t_k v_k : t_k\in [\underline{u}_k(x),\overline{u}_k(x)]\right\}.
\]
You can see existence of set-valued utility in the following theorem ~\ref{thm:existence_countable}:

\begin{theorem}[Existence for Countable \( X \)]
\label{thm:existence_countable}
Under Axioms \ref{axiom:interval_decomposition}--\ref{axiom:separability}, the map
\[
U(x) = \bigoplus_{k=1}^m I_k(x) v_k \in \mathcal{K}(\mathbb{R}^m),
\]
where \( v_1, \dots, v_m \) are the basis vectors in \( \mathbb{R}^m \) (i.e $\mathbb{R}^m= span\{v_1,...,v_m\}$), defines a zonotope-valued utility function for \( \pref \) such that
\[
x \pref y \quad \Longleftrightarrow \quad U(x) \ominus_e U(y) \subseteq \mathbb{R}^m_{\ge 0}.
\]
\end{theorem}
This Minkowski summation aggregates preferences across multiple criteria or dimensions, where each dimension represents an underlying interval order (criterion). Importantly, the linearity and convexity of Minkowski summation preserve key desirable properties needed for economic modeling, such as monotonicity and continuity.
\begin{corollary}[Minimal Dimension]
\label{cor:minimal_dimension}
The smallest \( m \) for which Theorem \ref{thm:existence_countable} holds is the interval dimension \( \dim_{\mathrm{int}}(X, \pref) \), defined as the smallest number of interval orders whose intersection equals \( \pref \).
\end{corollary}

The theorem establishes that \( U(x) \), constructed as the Minkowski sum of intervals along the standard basis, yields a zonotope in \( \mathcal{K}(\mathbb{R}^m) \). The corollary links the dimensionality of the representation to the intrinsic complexity of the preference relation. A critical feature of this construction is the role played by the choice of basis vectors $v_k$. The vectors $v_k$ determine how each interval order (criterion) contributes to the overall utility. In economic terms, these vectors represent \emph{weights} or directions along which each dimension of preference is evaluated. Component-wise positivity of the basis vectors, i.e., each $v_k$ having strictly nonnegative entries, ensures that the preferences are directionally consistent across criteria. The choice of different positive basis vectors reflects different importance or scaling of the underlying criteria, and consequently yields zonotopes with distinct shapes and combinational properties.

\begin{remark}[Overcomplete Generators]
\label{rem:overcomplete_generators}
In our construction we take \(m\) basis vectors \(v_1,\dots,v_m\) in \(\R^m\), one for each interval‐order component.  However, in general a zonotope in an ambient space \(\R^n\) may be generated by more segments than the dimension of that space—that is, one can have an \emph{overcomplete} set of generators. Concretely, if
\[
Z \;=\;\bigoplus_{k=1}^m I_k\,v_k
\qquad(v_k\in\R^n,\;m>n),
\]
then
\(\dim(Z)=\dim\bigl(\mathrm{span}\{v_1,\dots,v_m\}\bigr)\le n\),
and nothing in the Minkowski‐sum construction prevents us from choosing \(m>n\). In our representation theorem we simply embed into the “minimal’’ space \(\R^m\), but one could equally project those generators into a lower‐dimensional \(\R^n\) (with \(n< m\)) while preserving the full span.
\end{remark}

\subsection{The Uncountable Case}
\label{subsec:uncountable_case}

\subsubsection{Axioms}
\label{subsubsec:axioms_uncountable}

For an uncountable set \( X \), we retain Axioms \ref{axiom:interval_decomposition}--\ref{axiom:strict_consistency} and replace Axiom \ref{axiom:separability} with a stronger condition to accommodate the challenges of numerical representation in uncountable domains:

\begin{enumerate}[label=(A\arabic*), resume]
\item \label{axiom:separability} \emph{Separability}: For any \( x \spref y \), there exists \( z \in X \) such that \( x \pref z \pref y \).
\item \label{axiom:numerical_representability} \emph{Numerical Representability}: Each interval order \( \pref_k \) admits functions \( \underline{u}_k, \overline{u}_k: X \to \mathbb{R} \) such that
 \[
 x \pref_k y \iff \underline{u}_k(x) \ge \overline{u}_k(y), \quad \underline{u}_k(x) \le \overline{u}_k(x),
 \]
 and for any \( x \spref_k y \), there exists \( z \in X \) such that \( x \pref_k z \pref_k y \).
\end{enumerate}

Axiom \ref{axiom:numerical_representability} explicitly requires that each interval order has a numerical representation and retains the separability property, ensuring that the preference structure is rich enough for a zonotope-valued utility in uncountable settings.

\begin{corollary}\label{cor:convcomp}
    The $U$ as defined in Theorem \ref{thm:existence_countable} is compact and convex.
\end{corollary}

In many economic applications, we also require that the utility‐mapping be topologically well–behaved. If $X$ carries a natural topology and each interval‐order component admits continuous endpoints, then the zonotope‐valued map $U:X\to\mathcal K(\R^m)$ is itself continuous (where $\mathcal K(\R^m)$ is endowed with the Hausdorff metric). The following proposition ~\ref{prop:continuity} makes this precise.

\begin{proposition}[Continuity of the Zonotope‐Valued Utility]
\label{prop:continuity}
Let \(X\) be a Hausdorff topological space, and suppose \(\pref\) admits an interval‐order decomposition
\[
\pref \;=\;\bigcap_{k=1}^m\pref_k
\]
with each \(\pref_k\) represented by continuous endpoint functions
\[
\underline u_k,\;\overline u_k : X \;\longrightarrow\;\R,
\]
so that
\[
x\pref_k y \;\Longleftrightarrow\;\underline u_k(x)\ge\overline u_k(y),
\qquad
\underline u_k(x)\;\le\;\overline u_k(x)
\;\;\forall\,x\in X.
\]
Fix basis vectors \(v_1,\dots,v_m\in\R^m\) with nonnegative components, and define
\[
U(x)\;=\;\bigoplus_{k=1}^m\bigl[\underline u_k(x),\,\overline u_k(x)\bigr]\,v_k
\;\in\;\mathcal K(\R^m),
\]
where \(\mathcal K(\R^m)\) is endowed with the Hausdorff metric \(d_H\). Then
\[
U : X \;\longrightarrow\;\bigl(\mathcal K(\R^m),\,d_H\bigr)
\]
is continuous.
\end{proposition}

\paragraph{Product Space and Preference:} The space $\mathbb{R}^m$ is a product space, with each coordinate axis corresponding to an interval order $\pref_k$. The preference relation is defined via the extended set difference: $x \pref y \iff U(x) \ominus_e U(y) \subseteq \mathbb{R}^m_{\ge 0}$. Since $U(x), U(y)$ are zonotopes, $U(x) \ominus_e U(y) \subset \mathcal{K}(\mathbb{R}^m)$ is non-empty (Theorem \ref{thm:representation}). The condition $U(x) \ominus_e U(y) \subseteq \mathbb{R}^m_{\ge 0}$ aligns with the product order (Definition \ref{def:product-order}), where $a \ge b \iff a_i \ge b_i \forall i$, as $\mathbb{R}^m_{\ge 0}$ represents componentwise dominance. The basis vectors $v_k$ (e.g., $e_k$) ensure that $I_k(x)$ contributes to the $k$-th coordinate, preserving the independence of criteria, while non-standard positive bases allow weighted criteria (Section \ref{sec:examples}).

\subsection{Illustrative Examples of Zonotopes and Economic Interpretation}\label{sec:examples}

To clarify the geometric intuition behind zonotope-valued utilities and the economic meaning of basis vector selection, we now provide illustrative examples. We emphasize that different bases yield different zonotope shapes, thus encoding different patterns of preference aggregation and weighting.

\begin{example}[Two-dimensional Zonotope,Figure \ref{fig:basis_change_example}]
Consider two interval orders on $X$ represented by intervals $I_1(x)$ and $I_2(x)$. First, suppose the basis vectors are standard coordinate vectors: $v_1=(1,0)$ and $v_2=(0,1)$. The resulting zonotope for each alternative $x$ is a simple rectangle:
\[
U(x) = I_1(x)(1,0) + I_2(x)(0,1).
\]
If we instead select a nontrivial basis, such as $v'_1=(1,1)$ and $v'_2=(1,2)$, the zonotope transforms into a parallelogram oriented differently, reflecting an economically distinct weighting: here, the second criterion is twice as important vertically compared to the horizontal dimension.
\end{example}

\begin{figure}[ht!]
\centering
\includegraphics[width=0.5\textwidth]{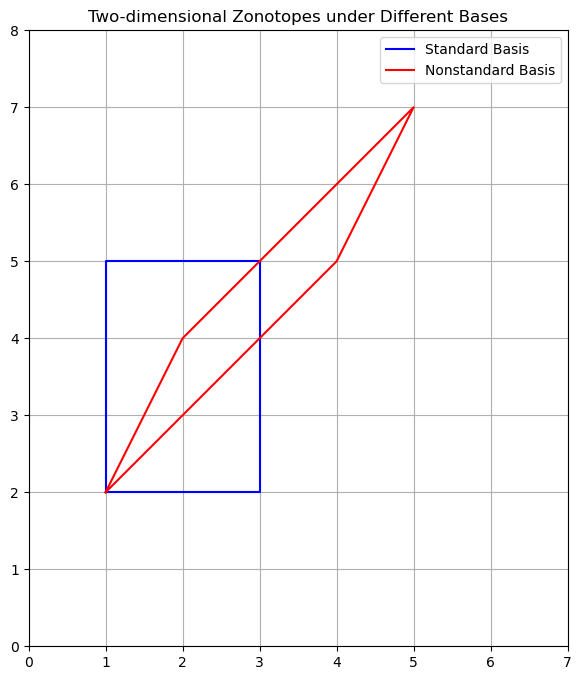}
\caption{Zonotopes for two interval orders under different bases: standard vs. nonstandard positive basis. The shape encodes relative weighting and thus the economic interpretation of underlying preferences.}
 \label{fig:basis_change_example}
\end{figure}

\begin{example}[Three-dimensional Zonotope, Figure \ref{fig:zonotope3d}]
Consider three interval orders corresponding to distinct economic criteria—for instance, price, quality, and sustainability. Suppose intervals reflect uncertainty or incompleteness in consumer preferences across these criteria. With basis vectors chosen as the standard positive orthonormal vectors $e_1, e_2, e_3$, each zonotope is a rectangular prism. However, if we alter basis vectors positively—such as $v_1=(1,1,0)$, $v_2=(0,1,1)$, and $v_3=(1,0,1)$—the resulting zonotope becomes a complex polyhedron (a rhombic dodecahedron). This geometrically encodes trade-offs and interaction effects between the criteria: each dimension no longer independently influences the overall preference but interacts with others, mirroring realistic decision-making contexts involving multidimensional trade-offs.
\end{example}

\begin{figure}[ht!]
 \centering
\includegraphics[width=0.5\textwidth]{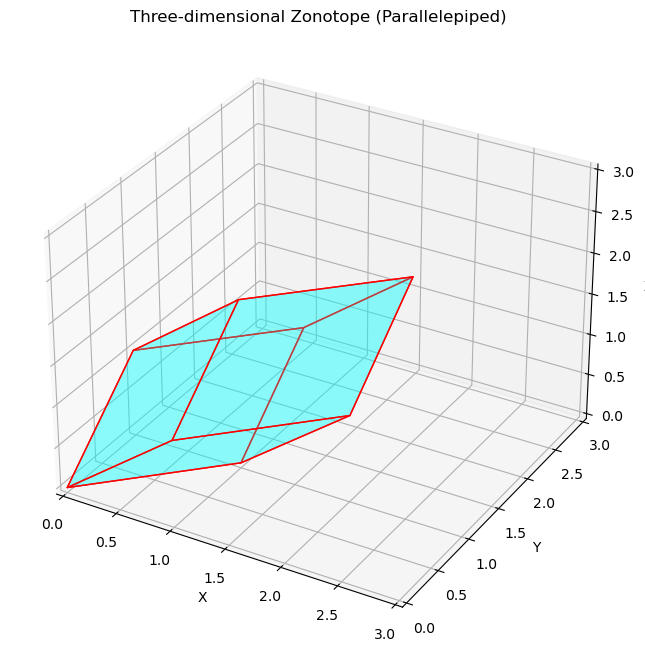}
\caption{Three-dimensional zonotope from nontrivial positive basis vectors. Economically, each vector direction represents a distinct weighting combination of economic attributes, which explicitly encodes multidimensional trade-offs and interactions among different attributes.}
\label{fig:zonotope3d}
\end{figure}
\par
\begin{remark}
These examples illustrate that the choice of basis vectors is economically significant, allowing practitioners to explicitly model how decision-makers perceive and weigh various attributes. Different positive bases yield zonotopes reflecting different trade-offs and aggregations of criteria, thus offering flexibility in practical applications of preference modeling.
\end{remark}

\par

The aggregation of intervals \( I_k(x) \) into the zonotope-valued utility \( U(x) \) via Minkowski summation is mathematically robust and conceptually justified. Minkowski summation is affine-linear, satisfying properties such as associativity, \(A\oplus(B\oplus C)=(A\oplus B)\oplus C\), and distributivity with nonnegative scalar multiplication, ensuring each interval contributes independently and proportionally to the representation. This linear structure guarantees that interval ordering is preserved: if \( I_k(x)\subseteq I_k(y) \), then \( U(x)\subseteq U(y) \), thus maintaining monotonicity crucial for preference consistency. Moreover, Minkowski summation inherently yields compact and convex zonotopes from compact intervals, ensuring each utility set \( U(x) \) is well-defined in the compact-convex family \( \mathcal{K}(\mathbb{R}^m) \). Philosophically, Minkowski summation captures the cumulative uncertainty or indifference inherent in multidimensional preference structures by combining all permissible points within each interval. Unlike scalar aggregation methods that collapse preferences to a single dimension, Minkowski summation preserves the multidimensional character of incomplete preferences, explicitly reflecting incomparabilities and partial orders through the geometric structure of the resulting zonotopes. The face lattice of these zonotopes intuitively encodes the logical structure of the underlying preferences, with facets corresponding to clear separations and lower-dimensional faces highlighting overlaps or incomparabilities. Therefore, Minkowski summation provides both a mathematically rigorous and philosophically meaningful approach, making zonotope-valued utilities an intuitively appealing and computationally effective representation of complex, incomplete preference relations.
\subsection{Representation Theorem}
\label{subsec:representation_theorems}

Now we summarize our development by unifying the axiomatic foundations, the existence theorems~\ref{thm:existence_countable}, and the minimality result of Corollary~\ref{cor:minimal_dimension} into a single representation statement. This theorem shows that, under the appropriate axioms (see Sections ~\ref{subsubsec:axioms_countable} and ~\ref{subsubsec:axioms_uncountable}), the preorder \(\pref\) is exactly captured by directional separations of the zonotope‑valued utility \(U\) via the extended set difference \(\ominus_{e}\) (see \citep{beresteanu2025extendedsetdifference} for more details on the extended set difference).
\begin{theorem}[Representation Theorem for Zonotope-Valued Utilities]
\label{thm:representation}
Let \(X\) be a set of alternatives (countable or uncountable) and \(\pref\) a preorder on \(X\). There exists a zonotope-valued utility function \(U: X \to \mathcal{K}(\mathbb{R}^m)\), where each \(U(x)\) is a zonotope, such that for all \(x, y \in X\),
\[
x \pref y \quad \Longleftrightarrow \quad U(x) \ominus_e U(y) \subseteq \mathbb{R}^m_{\ge 0},
\]
if and only if \(\pref\) satisfies the axioms outlined in Sections \ref{subsubsec:axioms_countable} for countable \(X\) or \ref{subsubsec:axioms_uncountable} for uncountable \(X\). Furthermore, the zonotope-valued utility function is defined as
\[
U(x) = \bigoplus_{k=1}^m I_k(x) v_k, \quad I_k(x) = [\underline{u}_k(x), \overline{u}_k(x)],
\]
where \(v_1, \dots, v_m\) are basis vectors in \(\mathbb{R}^m\) with strictly nonnegative components, and the dimension \(m\) is the interval dimension \(\dim_{\mathrm{int}}(X, \pref)\), as defined in Corollary \ref{cor:minimal_dimension}.
\end{theorem}
\begin{proposition}[Strict Preference Characterization]
\label{prop:strict_preference}
For alternatives \(x, y \in X\), \(x\) is strictly preferred to \(y\) (i.e., \(x \spref y\), meaning \(x \pref y\) and not \(y \pref x\)) if and only if either of the following equivalent conditions holds:
\begin{enumerate}
 \item The zonotope-valued utility \(U(x)\) lies strictly to the northeast of \(U(y)\), meaning that for every point \(u \in U(y)\), there exists a point \(v \in U(x)\) such that \(v \ge u\) (component-wise) with at least one strict inequality (i.e., \(v_i > u_i\) for some \(i\)).
 \item By a hyperplane separation theorem for compact convex sets in \(\mathbb{R}^m\), there exists a hyperplane defined by a normal vector \(w \in \mathbb{R}^m\) with strictly positive components (i.e., \(w_i > 0\) for all \(i\)) and a scalar \(c \in \mathbb{R}\) such that \(\sup_{u \in U(y)} \langle w, u \rangle \le c \le \inf_{v \in U(x)} \langle w, v \rangle\), with at least one strict inequality (i.e., \(\sup_{u \in U(y)} \langle w, u \rangle < c\) or \(c < \inf_{v \in U(x)} \langle w, v \rangle\)). This ensures that \(U(x)\) lies strictly on the northeast side (above) the hyperplane and \(U(y)\) on the southwest side (below).
\end{enumerate}
This characterization captures the directional dominance of \(U(x)\) over \(U(y)\), aligning with the extended set difference \(\ominus_e\) in Theorem \ref{thm:representation}, where strict preference corresponds to a clear separation in the positive orthant.
\end{proposition}

To connect the strict preference condition to the geometric properties of the extended set difference, we now present a proposition that explicitly links hyperplane separation with the position of \(U(x) \ominus_e U(y)\) in the positive orthant, providing a rigorous geometric interpretation of preference dominance.

\begin{proposition}[Hyperplane Separation and Extended Set Difference]
\label{prop:hyperplane_separation}
Let \(U(x), U(y) \in \mathcal{K}(\mathbb{R}^m)\) be zonotope-valued utilities for alternatives \(x, y \in X\). If there exists a hyperplane defined by a normal vector \(w \in \mathbb{R}^m\) with strictly positive components (i.e., \(w_i > 0\) for all \(i\)) and a scalar \(c \in \mathbb{R}\) such that \(\sup_{u \in U(y)} \langle w, u \rangle \le c \le \inf_{v \in U(x)} \langle w, v \rangle\), with at least one strict inequality (i.e., \(\sup_{u \in U(y)} \langle w, u \rangle < c\) or \(c < \inf_{v \in U(x)} \langle w, v \rangle\)), ensuring that \(U(x)\) lies strictly on the northeast side (above) the hyperplane and \(U(y)\) on the southwest side (below), then the extended set difference \(U(x) \ominus_e U(y)\), lies strictly in the positive orthant \(\mathbb{R}^m_{>0}\). That is, for any \(Z \in U(x) \ominus_e U(y)\), every point \(z \in Z\) satisfies \(z_i > 0\) for all \(i = 1, \dots, m\).
\end{proposition}
\begin{example}[Hyperplane Separation of Two Zonotope‐Valued Utilities]
\label{ex:hyperplane_separation}
Let us model two alternatives \(A\) and \(B\) by simple zonotopes in \(\R^2\), where each axis represents the utility of one attribute.

\begin{itemize}
 \item \textbf{Option \(A\):} For attribute 1 we allow any value in \([0,1]\), and for attribute 2 any value in \([0,0.5]\). Equivalently,
 \[
U(A)
= [0,1]\,(1,0)\;+\;[0,0.5]\,(0,1),
 \]
which is the parallelogram with vertices \(\{(0,0),(1,0),(1,0.5),(0,0.5)\}.\)

 \item \textbf{Option \(B\):} We take three generators of unit length in directions \((1,0)\), \((0.5,0.5)\) and \((0,1)\), then shift the whole shape by \((2,2)\). In formulas,
 \[
 U(B)
 = (2,2)\;+\;
[0,1]\,(1,0)\;+\;[0,1]\,(0.5,0.5)\;+\;[0,1]\,(0,1),
 \]
 which is a hexagon based at \((2,2)\).
\end{itemize}

\begin{figure}[ht]
 \centering
 \includegraphics[width=0.5\textwidth]{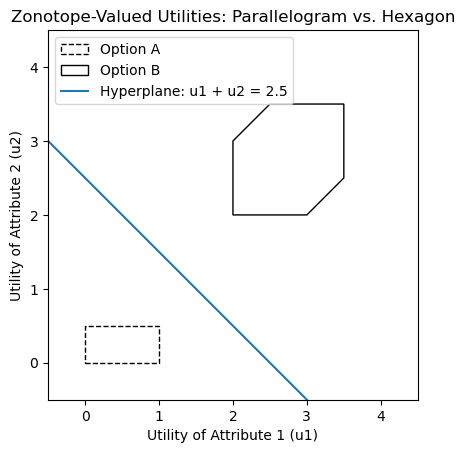}
 \caption{Parallelogram \(U(A)\) vs.\ hexagon \(U(B)\) in \(\R^2\), separated by the line \(u_1+u_2=2.5\).}
 \label{fig:hyperplane_separation_example}
\end{figure}

Now define the normal vector and threshold
\[
 w = (1,1),
 \qquad
 c = 2.5.
\]
A quick check shows
\[
 \sup_{u\in U(A)} w\!\cdot\!u
 = \max_{(u_1,u_2)\in U(A)}(u_1+u_2)
 = 1 + 0.5 = 1.5 < 2.5,
 \quad
 \inf_{v\in U(B)} w\!\cdot\!v
 = 2 + 2 = 4 > 2.5.
\]
Hence every point of \(U(A)\) lies strictly on the “south–west” side of the line \(u_1+u_2=c\), and every point of \(U(B)\) lies strictly on its “north–east” side.  By Proposition~\ref{prop:strict_preference}, this hyperplane separation implies
\[
  B \;\spref\; A,
\]
i.e.\ alternative \(B\) is strictly preferred to \(A\).
\end{example}
\begin{example}[Incomparable Alternatives via Overlapping Zonotopes]
\label{ex:incomparability}
Let \(X=\{A,B\}\) and represent their utilities in \(\R^2\) by

\[
 U(A)
 \;=\;
 [0,1]\,(1,0)
 \;+\;
 [0,1]\,(0,1)
 \;=\;\{(u_1,u_2):u_1,u_2\in[0,1]\},
\]
a unit square, and
\[
 U(B)
 \;=\;
 (0.5,0.5)
 \;+\;
[0,1]\,(1,0.2)
 \;+\;
 [0,1]\,(0.2,1),
\]
a parallelogram based at \((0.5,0.5)\).  One checks that
\[
 U(A)\;\cap\;U(B)
 =\{(u_1,u_2)\colon u_1,u_2\in[0.5,1]\}
 \;\neq\;\emptyset.
\]

\begin{figure}[ht]
\centering
\includegraphics[width=0.5\textwidth]{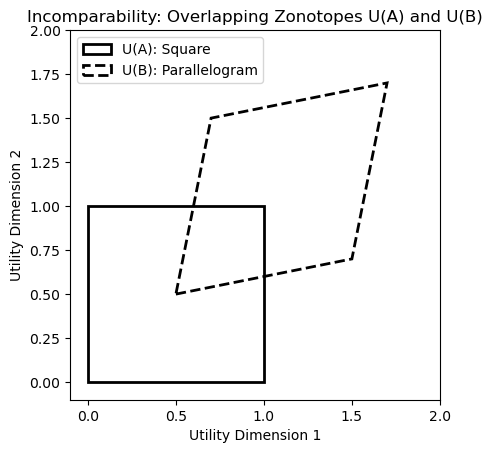}
\caption{Overlapping zonotopes \(U(A)\) (square) and \(U(B)\) (parallelogram).}
\label{fig:incomparability_example}
\end{figure}

Because \(U(A)\) and \(U(B)\) are not disjoint, no affine hyperplane can strictly separate them.  Equivalently, neither
\[
 U(A)\ominus_e U(B)\;\subseteq\;\R^2_{\ge0}
\quad\text{nor}\quad
 U(B)\ominus_e U(A)\;\subseteq\;\R^2_{\ge0}
\]
holds — each extended difference contains vectors with negative components. Geometrically this lack of separation corresponds exactly to the incomparability of \(A\) and \(B\) under \(\pref\).
\end{example}

\begin{remark}
The representation theorem, together with Propositions \ref{prop:strict_preference} and \ref{prop:hyperplane_separation}, provides a characterization of preferences and strict preferences representable by zonotope-valued utilities, extending the existence results from Theorems \ref{thm:existence_countable} and \ref{prop:continuity}. Proposition \ref{prop:strict_preference} formalizes strict preference through northeast separation or hyperplane separation with supremum and infimum conditions, ensuring that \(U(x)\) dominates \(U(y)\) in a directional sense. Proposition \ref{prop:hyperplane_separation} further clarifies that such hyperplane separation implies that the extended set difference \(\ominus_e\) yields a zonotope strictly in the positive orthant, reinforcing the geometric interpretation of preference dominance. The choice of basis vectors \(v_k\) with nonnegative components allows flexibility in modeling economic trade-offs, as illustrated in Section \ref{sec:zonotope_representation}, while the Minkowski summation ensures that the utility sets \(U(x)\) are compact and convex, preserving key properties like monotonicity and continuity essential for economic modeling.
\end{remark}

% --- Bibliography commands (for traditional BibTeX with natbib) ---
\bibliographystyle{ecta} % Use the Econometrica style BibTeX file (ecta.bst)
\bibliography{references}  % Your .bib file name (without the .bib extension)

\end{document}